\documentclass{JHEP3}
\usepackage{braket}
\usepackage{lineno}
\pdfoutput=1
\title { Formulae for the Analysis of the Flavor-Tagged Decay \boldmath{$B^0_s\rightarrow J/\psi \phi$}} 
\author{F.~Azfar$^{a}$, J.~Boudreau$^{b}$, N.~Bousson$^{c}$, J.~P.~Fern\'andez$^{d}$, K.~Gibson$^{b}$, G.~Giurgiu$^{e}$, G.~G\'omez-Ceballos$^{f}$, T.~Kuhr$^{h}$, M.~Kreps$^{h}$, C.~Liu$^{b}$, P.~Maksimovic$^{e}$, J.~Morlock$^{h}$,  L.~Oakes$^{a}$, M.~Paulini$^{g}$, E.~Pueschel$^{g}$, A.~Schmidt$^{h}$   \\
$^{a}$University of Oxford, Oxford OX1 3RH, United Kingdom \\ 
$^{b}$University of Pittsburgh,  Pittsburgh, PA 15260, U.S.A. \\  
$^{c}$Centre de Physique des Particules de Marseille, 12288 Marseille France\\
$^{d}$Centro de Investigaciones Energeticas Medioambientales y Tecnologicas, E-28040 Madrid, Spain \\
$^{e}$The John Hopkins University, Baltimore, MD 21218,U.S.A. \\
$^{f}$Massachusetts Institute of Technology, Cambridge, Massachusetts 02139, U.S.A. \\
$^{g}$Carnegie Mellon University, Pittsburgh, PA 15213, U.S.A. \\
$^{h}$Instit\"{u}t fur Experimentelle Kernphysik, Universit\"{a}t Karslruhe, 76128 Karlsruhe, Germany\\
Email:  \email{azfar@fnal.gov}, \email{boudreau@pitt.edu},  \email{bousson@cppm.in2p3.fr}, \email{fernand@fnal.gov},  \email{krg20@pitt.edu}, \email{ggiurgiu@jhu.edu}, \email{guillelmo.gomez-ceballos@cern.ch},  \email{thomas.kuhr@ekp.uni-karlsruhe.de}, \email{kreps@ekp.uni-karlsruhe.de}, \email{chl56@cmu.edu}, \email{petar@jhu.edu}, \email{morlock@ekp.uni-karlsruhe.de},  \email{loakes@fnal.gov}, \email{paulini@cmu.edu}, \email{epuesche@andrew.cmu.edu}, \email{aschmidt@ekp.uni-karlsruhe.de} }      
\abstract{
Differential rates in the decay $B^0_s \rightarrow J/\psi \phi$ with
$\phi\rightarrow K^+K^-$ and $J/\psi \rightarrow \mu^+\mu^-$ are
sensitive to the $CP$-violation phase
$\beta_s=\arg\left((-{V_{ts}V_{tb}^*) / (V_{cs}V_{cb}^*})\right)$,
predicted to be very small in the standard model.  The
analysis of $B^0_s \rightarrow J/\psi\phi$ decays is also suitable for
measuring the $B^0_s$ lifetime, the decay width difference $\Delta\Gamma_s$ 
between the $B^0_s$ mass eigenstates, and the $B^0_s$ oscillation frequency 
$\Delta m$ even if appreciable $CP$ violation does not occur. In this paper we present
normalized probability densities useful in maximum likelihood fits,
extended to allow for $S$-wave $K^+K^-$ contributions on one hand and for 
direct $CP$ violation on the other.  Our treatment of
the $S$-wave contributions includes the strong variation of the $S$-wave/$P$-wave
amplitude ratio with $m(K^+K^-)$ across the $\phi$ resonance, which was not considered in previous
work.  We include a scheme for re-normalizing the probability
densities after detector sculpting of the angular distributions of the
final state particles, and conclude with an examination of the
symmetries of the rate formulae, with and without an $S$-wave $K^+K^-$
contribution.  All results are obtained with the use of a new compact
formalism describing the differential decay rate of $B^0_s$ mesons
into $J/\psi \phi$ final states.
}


\preprint

\begin{document}
%\begin{titlepage}

\setcounter{page}{0}
\thispagestyle{empty}
%\end{titlepage}

%\tableofcontents
%\listoffigures
%\listoftables
\pagenumbering{arabic}
%\linenumbers

\section{Introduction}

The decay of $B^0_s \rightarrow J/\psi \phi$, a transition of a pseudoscalar into
two vector mesons can be thought of as six
independent decays. The initial $B^0_s$ system consists
of a heavy and a light mass eigenstate, and the $J/\psi \phi$ system to which
it decays is
characterized by three distinct orbital angular momentum states.  A maximum
amount of information about this system can be obtained from analyses which
disentangle the two initial states and the three final states.  The
experimental technique of \emph{flavor tagging} infers a meson's flavor at
production time as $B^0_s$ or ${\bar B}^0_s$ and is 
the key to disentangling the two initial
states. Flavor-tagged $B^0_s \rightarrow J/\psi \phi$ decays are of
great interest in particle physics because of their sensitivity to the
CKM phases~\cite{ref:digheCKM} and to anomalous mixing phases from
physics beyond the standard model~\cite{ref:LenzAndNierste}.  Recently
the CDF and D0 collaborations have constrained the CKM phases in both
untagged analyses~\cite{ref:CDFUntagged,ref:D0Untagged}, and
flavor-tagged analyses~\cite{ref:CDFTagged,ref:D0Tagged}.
%, the latter
%having been recently updated~\cite{ref:CdfUnpublished,ref:D0Unpublished}. 
These  analyses are based on complete differential rates for the decay given 
in Ref.~\cite{ref:InPursuit}. They use the angular 
distributions of the decay products to disentangle the three final states.

In this paper we re-express the differential decay rates in Ref.~\cite{ref:InPursuit}, 
using a new formalism that makes explicit a number of symmetries that are 
otherwise hidden.  These formulae are then extended to the case in which the 
final state in the decay  $B^0_s \rightarrow J/\psi \phi$ includes decays 
of type $B^0_s \rightarrow J/\psi K^+ K^-$ (kaons in an $S$-wave state), which 
has been suggested~\cite{ref:Sheldon} to be an important effect.  After including the 
$S$-wave contribution in the theoretical description, we identify the symmetries of the 
modified formulae. 

In addition to $S$-wave effects, we also investigate other aspects of the 
differential decay rate formulae.  We include the effects of possible direct 
$CP$ violation.  In addition we show how interference between $CP$ odd and 
$CP$ even  $J/\psi \phi$ final states effectively tags the flavor of the $B^0_s$ meson at
decay, allowing for the possibility to observe $B^0_s \rightarrow \bar {B}^0_s$ 
flavor oscillations in a flavor-tagged analysis, even in the absence 
of $CP$ violation effects. 

Experimentally, the differential rate formulae are used to construct likelihood
functions based on normalized probability density functions (PDFs). In this
paper we include normalization constants where appropriate in all expressions
for transition amplitudes and PDFs. Detector angular acceptance is an important 
effect which must be included in these probability densities.
However, the inclusion of this effect disturbs the normalization of the PDF.  We 
present a scheme for normalizing the probability density analytically, as required
for unbinned maximum likelihood fits. 

\section {Phenomenology of the $B^0_s \rightarrow J/\psi \phi$ Decay}

We first summarize the phenomenology of the $B^0_s$ system and the decay $B^0_s \rightarrow J/\psi \phi \rightarrow \mu^+ \mu^- K^+ K^-$\,.
Two flavor eigenstates, $\ket{B^0_s}$ and $\ket{{\bar B}^0_s}$, mix via the weak interaction.  The two mass eigenstates 
% ++++++++++++++++++++++++++++++++++++++++++++++++++++++++++++++++++++++
\begin{displaymath}
|B_s^H\rangle = p\,|B_s^0\rangle - q\,|\bar B_s^0\rangle,
%\hspace*{0.9cm} 
\qquad
|B_s^L\rangle = p\,|B_s^0\rangle + q\,|\bar B_s^0\rangle
\end{displaymath}
% ++++++++++++++++++++++++++++++++++++++++++++++++++++++++++++++++++++++
are labeled ``heavy'' and ``light''.  The mass and lifetime differences 
between the $B_s^H$ and $B_s^L$ states can be defined as
% ++++++++++++++++++++++++++++++++++++++++++++++++++++++++++++++++++++++
\begin{displaymath}
\Delta m \equiv m_H - m_L,\ \ \ \Delta\Gamma \equiv \Gamma_L - \Gamma_H,\ \ 
\ \ \Gamma = \,(\Gamma_H + \Gamma_L)/2\,, \
\end{displaymath}
% ++++++++++++++++++++++++++++++++++++++++++++++++++++++++++++++++++++++
where $m_{H,L}$ and $\Gamma_{H,L}$ denote the mass and decay width of
$B_s^H$ and $B_s^L$ (with this definition both $\Delta m$ and
$\Delta\Gamma$ are expected to be positive quantities). The heavy state decays with a longer lifetime,  $\tau_H= 1/\Gamma_H$,
while the light state decays with the shorter lifetime $\tau_L= 1/\Gamma_L$, in analogy to the neutral kaon system. 
The mean lifetime is defined to be $\tau= 1/\Gamma$. Theoretical estimates predict $\Delta\Gamma/\Gamma$ 
to be on the order of $\sim15\%$ \cite{ref:LenzAndNierste}. Linear polarization eigenstates of the $J/\psi$ and $\phi$ 
provide a convenient basis for the analysis of the decay~\cite{ref:dighe}.  The two vector mesons can have their spins transversely polarized with
respect to their momentum and be either parallel or perpendicular to each other.  Alternatively, they can
both be longitudinally polarized. We denote these states as $\ket{{\cal P}_{||}}$,
$\ket{{\cal P}_{\perp}}$, and $\ket{{\cal P}_{0}}$.

In the standard model, $CP$ violation occurs through complex phases in
the CKM matrix~\cite{ref:CK}.  Large phases occur in the matrix elements $V_{ub}$
and $V_{td}$. While these matrix elements generate large $CP$ violation
in the $B^0$ system, they do not appear in leading order diagrams
contributing to either $B^0_s \leftrightarrow {\bar B^0_s}$ mixing or
to the decay $B^0_s\rightarrow J/\psi \phi$.  For this reason the
standard model expectation of $CP$ violation in $B^0_s \rightarrow
J/\psi \phi$ is small.  In the limit of vanishing $CP$ violation, the heavy, long-lived mass eigenstate $B_s^H$
is $CP$ odd and decays to the $CP$-odd, $L$=1 orbital angular momentum
state $\ket{{\cal P}_{\perp}}$.   The light, short-lived mass eigenstate $B_s^L$ is $CP$ even
and decays to both $CP$-even $L$=0 and $L$=2 orbital angular momentum states,
which are linear combinations of $\ket{{\cal P}_{0}}$ and $\ket{{\cal P}_{||}}$.

The small $CP$ violation in $B^0_s\rightarrow J/\psi \phi$ can be quantified in the 
following way:  we define $A_i$ as the decay amplitude $\langle B_s|H|{\cal P}_i \rangle$ and 
${\bar A_i}$ as the decay amplitude  $\langle{\bar B_s}|H|{\cal P}_i\rangle$ where $i$ is one of
$\{||, \perp, 0\}$. All $CP$ observables in the system are characterized by three quantities $\lambda_i = \frac{q}{p} \frac {\bar A_i}{A_i}$.   In the standard model
the $\lambda_i$ are given as $\lambda_i = \pm\exp {\left (i2\beta_s \right)}$
where the positive and negative sign applies to the $CP$ even and odd final state, and 
\begin{displaymath}
\beta_s\equiv\arg\left(-{V_{ts}V_{tb}^*\over V_{cs}V_{cb}^*}\right).
\end{displaymath} 
The standard model 
expectation~\cite{ref:utfit} is $2\beta_s = $ 0.037 $\pm$ 0.002, a very small
phase which does not lead to appreciable levels of $CP$ violation.  New physics can alter the mixing phase,
while leaving $\lambda$ very nearly unimodular.   In this paper we consider, however, 
also the case in which $|\lambda| \ne 1$.

\section{Differential Rates}

The state of an initially pure $B^0_s$ or $\bar{B}^0_s$ meson after a proper time $t$ has elapsed is denoted as $|B^0_{s,phys}(t)\rangle$ and $|{\bar B}^0_{s,phys}(t)\rangle$. 
Transitions of these states to the detectable $\mu^+\mu^-K^+K^-$ can be written as  
\begin{eqnarray}
 \langle \mu^+\mu^- &K&^+K^-|H|B^0_{s,phys}(t) \rangle   \nonumber \\
\hspace{2cm}& = &\sum_{i}{\braket{\mu^+\mu^-K^+K^-|H|{\cal P}_i}  \braket{{\cal P}_i|H|B^0_s}        \braket{B^0_s|B^0_{s,phys}(t)}       }  \nonumber \\
\hspace{2cm}& + &\sum_{i}{\braket{\mu^+\mu^-K^+K^-|H|{\cal P}_i}  \braket{{\cal P}_i|H|\bar{B^0_s} } \braket{\bar{B^0_s}|B^0_{s,phys}(t)}     } , \nonumber \\
\langle\mu^+\mu^-&K&^+K^-|H|\bar{B}^0_{s,phys}(t)\rangle  \nonumber \\
\hspace{2cm}& = &\sum_{i}{\braket{\mu^+\mu^-K^+K^-|H|{\cal P}_i}  \braket{{\cal P}_i|H|B^0_s}        \braket{B^0_s|\bar{B}^0_{s,phys}(t)}       }  \nonumber \\
\hspace{2cm}& + &\sum_{i}{\braket{\mu^+\mu^-K^+K^-|H|{\cal P}_i}  \braket{{\cal P}_i|H|\bar{B^0_s} } \braket{\bar{B^0_s}|\bar{B}^0_{s,phys}(t)}     } . \nonumber \\
\label{eqn:startingPoint}
\end{eqnarray}
where $H$ is the weak interaction Hamiltonian.
The expression can be written much more simply, by defining time-dependent amplitudes for $\ket{B^0_s}$ and $\ket{{\bar B}^0_s}$ to reach the states $\ket{{\cal P}_i}$ either with or without mixing:
\begin{eqnarray}
{\mathcal A}_i(t)         & \equiv & \braket{{\cal P}_i|H|B^0_s}\braket{B^0_s|B^0_{s,phys}(t)}+\braket{{\cal P}_i|H|\bar{B^0_s}}\braket{\bar{B^0_s}|B^0_{s,phys}(t)} , \nonumber \\
%                      &     =  & {e^{-imt}e^{-\Gamma t/2}} \left (E_+(t) A_i + \frac {q}{p}E_-(t) {\bar A}_i \right ) \nonumber \\
{{\bar {\mathcal A}}}_i(t)  & \equiv & \braket{{\cal P}_i|H|B^0_s}\braket{B^0_s|{\bar B}^0_{s,phys}(t)}+\braket{{\cal P}_i|H|\bar{B^0_s}}\braket{\bar{B^0_s}|{\bar B}^0_{s,phys}(t)} . \nonumber 
%                      &     =  & {e^{-imt}e^{-\Gamma t/2}} \left (\frac {p}{q} E_-(t) A_i + E_+(t) {\bar A}_i \right )
\label{eqn:ADef}
\end{eqnarray}
Then:
\begin{eqnarray}
\braket{\mu^+\mu^-K^+K^-|H|B^0_{s,phys}(t)}         = \sum_{i}{{\mathcal A}_i(t)e^{-imt}}        \braket{\mu^+\mu^-K^+K^-|H|{\cal P}_i} ,  \nonumber \\
\braket{\mu^+\mu^-K^+K^-|H|{\bar B}^0_{s,phys}(t)}  = \sum_{i}{{\bar {\mathcal A}}_i(t)e^{-imt}} \braket{\mu^+\mu^-K^+K^-|H|{\cal P}_i} \,,  \nonumber \\
\label{eqn:simple0}
\end{eqnarray}
where the time dependence of ${\cal A}_i(t)$ and ${\bar {\cal A}_i}(t)$ is:
\begin{eqnarray}
{\cal A}_i(t) &=& \frac{e^{-\Gamma t /2}}{\sqrt{\tau_H + \tau_L \pm \cos{2\beta_s}\left(\tau_L-\tau_H\right)}} \left[
E_+(t) \pm e^{2i\beta_s} E_-(t)
\right] a_i\,,   \nonumber \\
{\bar {\cal A}_i}(t) &=& \frac{e^{-\Gamma t /2}}{\sqrt{\tau_H + \tau_L \pm \cos{2\beta_s}\left(\tau_L-\tau_H\right)}} \left[
\pm E_+(t) + e^{-2i\beta_s} E_-(t)
\right] a_i\,,  \nonumber \\
\label{eqn:finalAmp}
\end{eqnarray}
% ++++++++++++++++++++++++++++++++++++++++++++++++++++++++++++++++++++++
and where the upper sign indicates a $CP$~even final state, the lower sign indicates a $CP$~odd final state,
\begin{equation}
E_{\pm}(t) \equiv  \frac{1}{2}\left[e^{+\left(\frac{-\Delta\Gamma}{4} + i\frac{\Delta m}{2}\right)t} \pm e^{-\left(\frac{-\Delta\Gamma}{4} + i\frac{\Delta m}{2}\right)t}\right],
\label{eqn:functionDef}
\end{equation}
and the $a_i$ are complex amplitude parameters satisfying:
\begin{equation}
\sum_i {|a_i|^2} = 1 \,.
\label{eqn:aNorm}
\end{equation}

%The mixing amplitudes are
%\begin{eqnarray}
%\braket{B^0_{s,phys}(t)|B^0_s}  =   {e^{-imt}e^{-\Gamma t/2}} E_+(t),             \ \ \  \braket{B^0_{s,phys}(t)|{\bar B}^0_s}  =  {e^{-imt}e^{-\Gamma t/2}} \frac {q}{p}E_-(t)  \nonumber \\
%\braket{B^0_{s,phys}(t)|B^0_s}  =   {e^{-imt}e^{-\Gamma t/2}} \frac {p}{q}E_-(t), \ \ \  \braket{B^0_{s,phys}(t)|{\bar B}^0_s}  =  {e^{-imt}e^{-\Gamma t/2}}             E_+(t) 
%\end{eqnarray}
%where
 
The final state $\mu^+\mu^-K^+K^-$ is characterized by three decay angles, described in a 
coordinate system\footnote{An alternate basis called the helicity basis is 
discussed further in Section~\ref{sec:symmetries}.} called the transversity basis~\cite{ref:digheCKM}.  In the $J/\psi$ rest frame, 
the $x$-axis is taken to lie  along the momentum of the $\phi$ and the $z$-axis perpendicular to the decay 
plane of the $\phi$.  The variables ($\theta$, $\varphi$) are the polar and azimuthal angles of the $\mu^+$
momentum in this basis. We also define the angle $\psi$ to be the ``helicity'' angle in the $\phi$
decay, i.e. the angle between the $K^+$ direction  and the $x$-axis in the $\phi$ rest frame.
With these definitions, the muon momentum direction in the $J/\psi$ rest frame is given by the unit vector
% ++++++++++++++++++++++++++++++++++++++++++++++++++++++++++++++++++++++
\begin{equation}
\hat{n} = \left(\sin{\theta}\cos{\varphi}, \sin{\theta}\sin{\varphi}, \cos{\theta} \right).
\label{eqn:nhat}
\end{equation}
% ++++++++++++++++++++++++++++++++++++++++++++++++++++++++++++++++++++++
Let ${\bf A}(t)$ and ${\bf {\bar A}}(t)$ be complex vector functions of time defined as
% ++++++++++++++++++++++++++++++++++++++++++++++++++++++++++++++++++++++
\begin{eqnarray}
{\bf A}(t)=\left({\mathcal A}_0(t)\cos{\psi}, -\frac{{\mathcal A}_\parallel(t)\sin{\psi}}{\sqrt{2}}, i\frac{{\mathcal A}_\perp(t)\sin{\psi}}{\sqrt{2}}\right),  \nonumber \\
{\bf{\bar {A}}}(t)=\left({\bar {\mathcal A}}_0(t)\cos{\psi}, -\frac{{\bar {\mathcal A}}_\parallel(t)\sin{\psi}}{\sqrt{2}}, i\frac{{\bar {\mathcal A}}_\perp(t)\sin{\psi}}{\sqrt{2}}\right),
\label{eqn:fixedAngle}
\end{eqnarray}
where ${\cal A}_i(t)$ have now been normalized.
For experimental measurements we are concerned with normalized probability density functions $P_B$ and $P_{\bar B}$ for $B$ and ${\bar B}$ mesons
in the variables $t$, $\cos{\psi}$, $\cos{\theta}$, and $\varphi$, which can be obtained by squaring Eq.~(\ref{eqn:simple0}).   The formulae of Ref.~\cite{ref:InPursuit} are then equivalent to: 
\begin{eqnarray} 
P_{B}(\theta, \varphi, \psi, t) = \frac{9}{16\pi} |{\bf A}(t)\times \hat{n}|^2  \nonumber \\
P_{\bar{B}}(\theta, \varphi, \psi, t) = \frac{9}{16\pi} |{\bf {\bar A}(t)}\times \hat{n}|^2
\label{eqn:finalAngle}
\end{eqnarray}
% ++++++++++++++++++++++++++++++++++++++++++++++++++++++++++++++++++++++
which give a picture of a time-dependent polarization analyzed in the decay\footnote{Throughout this paper, when writing the dot product of two complex vectors, we always imply complex conjugation on the second operand.}.  The factors of 9/16$\pi$ are normalization constants, and are present in order that
% ++++++++++++++++++++++++++++++++++++++++++++++++++++++++++++++++++++++
\begin{equation}
\int \sum_{j=B,\bar{B}} P_j(\psi, \theta, \varphi, t) d(\cos{\psi}) d(\cos{\theta}) d\varphi dt = 1 \,.
\label{eqn:goodNorm}
\end{equation}
% ++++++++++++++++++++++++++++++++++++++++++++++++++++++++++++++++++++++
The quantities $|a_i|^2$ give the time-integrated rate to each of the polarization states.  The values of 
${\cal A}_i(t)$ at $t=0$ will be denoted as $A_i$.  To translate between the $a$'s and the $A$'s
one can use the following two sets of transformations:
\begin{eqnarray}
|A_{\perp}|^2 = \frac{|a_{\perp}|^2y}{1+(y-1)|a_{\perp}|^2}  \qquad    |a_{\perp}|^2 = \frac{|A_{\perp}|^2}{y+(1-y)|A_{\perp}|^2}                         \nonumber \\
|A_{||}|^2 = \frac{|a_{||}|^2}{1+(y-1)|a_{\perp}|^2}         \qquad    |a_{||}|^2 = \frac{|A_{||}|^2 y}{y+(1-y)|A_{\perp}|^2}                         \nonumber \\
|A_0|^2 = \frac{|a_0|^2}{1+(y-1)|a_{\perp}|^2}               \qquad    |a_0|^2 = \frac{|A_0|^2 y}{y+(1-y)|A_{\perp}|^2}                           \nonumber \\
\end{eqnarray}
where $y \equiv (1-z)/(1+z)$ and $z\equiv \cos{2\beta_s}\Delta\Gamma/(2\Gamma)$.  The relation~(\ref{eqn:aNorm}) insures that
\begin{equation}
\sum_i {|A_i|^2} = 1
\end{equation}
Eq.~(\ref{eqn:finalAngle}), together with the definitions in Eqs.~(\ref{eqn:finalAmp}), (\ref{eqn:functionDef}), and (\ref{eqn:nhat}) can be used as
a decay model for an event generator, and is suitable for use as a fitting function in the absence of detector effects.

\section{Detector Efficiency and Normalization} 
\label{chap:normalization}
% ++++++++++++++++++++++++++++++++++++++++++++++++++++++++++++++++++++++
The detector efficiency $\varepsilon(\psi,\theta, \varphi)$, when introduced into the above
expression, disturbs the normalization of Eq.~(\ref{eqn:goodNorm}).  We restore it by
dividing by a normalization factor $N$,
% ++++++++++++++++++++++++++++++++++++++++++++++++++++++++++++++++++++++
\begin{eqnarray}
  &P&^\prime(\psi, \theta, \varphi, t) = \frac{1}{N} P(\psi, \theta, \varphi, t) \varepsilon(\psi, \theta, \varphi)\,, \nonumber \\ 
  &N& = \int \sum_{i=B,\bar{B}} P_i(\psi, \theta, \varphi, t)  \varepsilon(\psi, \theta, \varphi) d(\cos{\psi}) d(\cos{\theta}) d\varphi dt\,. \nonumber \\
\label{eqn:normal1}
\end{eqnarray}
% ++++++++++++++++++++++++++++++++++++++++++++++++++++++++++++++++++++++
Suppose that the efficiency $\varepsilon(\psi, \theta, \varphi)$ can be parametrized
as 
% ++++++++++++++++++++++++++++++++++++++++++++++++++++++++++++++++++++++
\begin{equation}
  \varepsilon(\psi, \theta, \varphi) = c_{lm}^{k}P_k(\cos{\psi}) Y_{lm}(\theta, \varphi), \label{eqn:Pprimeprime}
\end{equation}
% ++++++++++++++++++++++++++++++++++++++++++++++++++++++++++++++++++++++
where $c_{lm}^k$ are expansion coefficients, $P_k(\cos{\psi})$ are Legendre
polynomials, and $Y_{lm}(\theta, \varphi)$ are real harmonics related to the
spherical harmonics through the following relations:
\begin{eqnarray}
Y_{lm}=& Y_l^m    & (m=0)\,, \nonumber \\
Y_{lm}=& \frac{1}{\sqrt{2}} ( Y_l^m + (-1)^{m}Y_l^{-m}) & (m >  0)\,, \nonumber \\
Y_{lm}=& \frac{1}{i\sqrt{2}} ( Y_l^{|m|} - (-1)^{|m|}Y_l^{-|m|}) & (m <  0)\,.\label{eqn:ylm}
\end{eqnarray}
The products $P_k(\cos{\psi}) Y_{lm}(\theta, \varphi)$ constitute an orthonormal basis for
functions  of the three angles. The detector efficiency (obtained, for example, from Monte Carlo simulation) can be 
fit to the first few of these polynomials.  A straight-forward calculation 
shows that:
% ++++++++++++++++++++++++++++++++++++++++++++++++++++++++++++++++++++++
\begin{eqnarray}
\label{eqn:normFactor}
 N  = & &\nonumber \\
&   &  \frac{3}{8\sqrt{ \pi}}   \left[   \frac{4c_{00}^0}{3}(|a_0|^2 + |a_{\parallel}|^2 + |a_{\perp}|^2) \right. \nonumber \\
& + &\left.\frac{4c_{00}^2}{15}(2|a_0|^2 - |a_{\parallel}|^2 - |a_{\perp}|^2)   \right]  \nonumber \\
    & + & \frac{3}{8\sqrt{5\pi}}   \left[   \frac{2c_{20}^0}{3}( |a_0|^2 + |a_{\parallel}|^2 - 2|a_{\perp}|^2) \right. \nonumber \\
& + & \left. \frac{4c_{20}^2}{15}(|a_0|^2 - \frac{1}{2}|a_{\parallel}|^2 + |a_{\perp}|^2)   \right]  \nonumber \\
    & - &  \frac{9}{16\sqrt{15\pi}}                   \frac{\sin{2\beta_s} (\tau_L-\tau_H)}{   \sqrt{ ((\tau_L - \tau_H)\sin{2\beta_s})^2   +{4\tau_L\tau_H} }} \nonumber \\
 & & \times  \left[(a_{\parallel}^*a_{\perp}+a_{\parallel}a_{\perp}^*)(\frac{4}{3} c_{2-1}^0 - \frac{4}{15} c_{2-1}^2)    \right] \nonumber \\
    & + &  \frac{9}{16} \frac{\sqrt{2}}{\sqrt{15\pi}} \frac{\sin{2\beta_s} (\tau_L-\tau_H)}{   \sqrt{ ((\tau_L - \tau_H)\sin{2\beta_s})^2   +{4\tau_L\tau_H} }} \nonumber \\ 
& & \times  \left[(a_{0}^*a_{\perp}+a_{0}a_{\perp}^* ) (\frac{\pi c_{21}^{1}}{8} - \frac{\pi c_{21}^3}{32} + ... ) \right] \nonumber \\ 
    & + & \frac{9}{8\sqrt{15\pi}}   \left[   \frac{2c_{22}^0}{3}(-|a_0|^2 +|a_{\parallel}|^2) - \frac{4c_{22}^2}{15}(|a_0|^2 + \frac{1}{2}|a_{\parallel}|^2)   \right]  \nonumber \\
    & + & \frac{9}{16} \frac{\sqrt{2}}{\sqrt{15\pi}}  \left[(a_{0}^*a_{\parallel}+a_{0}a_{\parallel}^*) (\frac{\pi c_{2-2}^{1}}{8} - \frac{\pi c_{2-2}^3}{32} + ... ) \right].
\end{eqnarray}
% ++++++++++++++++++++++++++++++++++++++++++++++++++++++++++++++++++++++
The numerical factors $+\pi/8$ and $-\pi/32$, appearing together with $c^k_{2,1}$ and $c^k_{2,-2}$ in the infinite series, 
are the integrals
\begin {equation}
\int P_k(\cos{\psi}) \cos (\psi) \sin{\psi} d(\cos{\psi}) \,.
\label{eqn:expansion}
\end{equation}
While this series is infinite, the number of basis functions needed to fit detector efficiencies in
a particular analysis is finite and determined chiefly by the size of the data sample.  With the
factors in Eq.~(\ref{eqn:expansion}) the normalizing factor can be adapted to account for all terms used in
the expansion of the efficiency. Eq.~(\ref{eqn:normFactor}) represents an analytic normalization of the fitting function and provides an
efficient way to compute the likelihood during a maximum log
likelihood fit. The orthonormality of the basis functions has been used to reduce the expression to its final form.

\section {Time Development}
\label{section:timeDev}
The short oscillation length of the $B^0_s$ meson~\cite{ref:CDFB0s}, $2\pi c/\Delta m\sim 106~\mu$m,
requires us to account for resolution effects when fitting the rates of flavor-tagged
decays, even using the best silicon vertex detectors, which have proper decay length resolutions on the order of 25~$\mu$m. 
Certain time-dependent functions arising from particle-antiparticle oscillations, particularly
those expressed as the product of exponential decays and harmonic functions with frequency
$\Delta m$, must be convolved with one or more Gaussian components describing detector resolution.
This convolution can be carried out analytically, using the method described in 
Ref.~\cite{ref:wwerf} for the evaluation of certain integrals which are equivalent 
to complex error functions.  In this step one requires that various components of the 
time dependence first be separated from Eq.~(\ref{eqn:finalAngle}).  The time development
of ${\cal A}_0(t)$ and ${\cal A}_{\parallel}(t)$ amplitudes are identical, 
but differs from that of ${\cal A}_{\perp}(t)$. We begin by decomposing
% ++++++++++++++++++++++++++++++++++++++++++++++++++++++++++++++++++++++
\begin{equation}
{\bf A} (t) =  {\bf A}_+(t) + {\bf A}_-(t), \ \ \ {\bf {\bar A}} (t) =  {\bf {\bar A}}_+(t) + {\bf {\bar A}}_-(t)
\end{equation}
% ++++++++++++++++++++++++++++++++++++++++++++++++++++++++++++++++++++++
where
% ++++++++++++++++++++++++++++++++++++++++++++++++++++++++++++++++++++++
\begin{eqnarray}
  {\bf A}_+(t)        = {\bf A}_+ f_+(t) = (a_0\cos{\psi}, -\frac{a_\parallel\sin{\psi}}{\sqrt{2}},0)\cdot f_+(t) \,,\nonumber \\
  {\bf {\bar A}}_+(t) = {\bf {\bar A} }_+ {\bar f}_+(t) = (a_0\cos{\psi}, -\frac{a_\parallel\sin{\psi}}{\sqrt{2}},0)\cdot {\bar f}_+(t)\,,
\end{eqnarray}
% ++++++++++++++++++++++++++++++++++++++++++++++++++++++++++++++++++++++
and
% ++++++++++++++++++++++++++++++++++++++++++++++++++++++++++++++++++++++
\begin{eqnarray}
{\bf A}_-(t)=  {\bf A}_- f_-(t) = (0,0, i\frac{a_\perp\sin{\psi}}{\sqrt{2}}) \cdot f_-(t) \,,\nonumber \\
{\bf {\bar A}}_-(t)=  {\bf {\bar A}}_- {\bar f}_-(t) = (0,0, i\frac{a_\perp\sin{\psi}}{\sqrt{2}}) \cdot {\bar f}_-(t)\,,
\end{eqnarray}
and we define 
\begin{eqnarray}
f_\pm(t) &=& \frac{e^{-\Gamma t /2}}{\sqrt{\tau_H + \tau_L \pm \cos{2\beta_s}\left(\tau_L-\tau_H\right)}} \left[
E_+(t) \pm e^{2i\beta_s} E_-(t)
\right] \,,   \nonumber \\
{\bar f_\pm}(t) &=& \frac{e^{-\Gamma t /2}}{\sqrt{\tau_H + \tau_L \pm \cos{2\beta_s}\left(\tau_L-\tau_H\right)}} \left[
\pm E_+(t) + e^{-2i\beta_s} E_-(t)
\right] \, .  \nonumber \\
\end{eqnarray}
% ++++++++++++++++++++++++++++++++++++++++++++++++++++++++++++++++++++++
We then have in place of Eq.~(\ref{eqn:finalAngle})
% ++++++++++++++++++++++++++++++++++++++++++++++++++++++++++++++++++++++
\begin{eqnarray}
& P_{B}& (\theta, \psi, \varphi, t)    \nonumber \\
& =  & \frac{9}{16\pi}  \left\{ |{\bf A}_+(t)\times \hat{n}|^2 +  |{\bf A}_-(t)\times \hat{n}|^2 + 2 Re(({\bf A}_+(t)\times \hat{n}) \cdot ({\bf A}_-^*(t)\times \hat{n})) \right\} \nonumber \\
& = &  \frac{9}{16\pi} \left\{ |{\bf A}_+\times \hat{n}|^2 |f_+(t)|^2 + |{\bf A}_-\times \hat{n}|^2 |f_-(t)|^2 \right. \nonumber \\ 
&  & +              \left. 2 Re(({\bf A}_+\times \hat{n}) \cdot ({\bf A}_-^*\times \hat{n})\cdot f_+(t)\cdot f_-^*(t))\right\} 
\label{eqn:fullProb1}
\end{eqnarray}
% ++++++++++++++++++++++++++++++++++++++++++++++++++++++++++++++++++++++
and
% ++++++++++++++++++++++++++++++++++++++++++++++++++++++++++++++++++++++
\begin{eqnarray}
&P_{\bar{B}}&(\theta, \psi, \varphi, t)   \nonumber \\
& = & \frac{9}{16\pi}  \left\{ |{\bf \bar{A}}_+(t)\times \hat{n}|^2 +  |{\bf \bar{A}}_-(t)\times \hat{n}|^2 + 2 Re({\bf \bar{A}}_+(t)\times \hat{n}) \cdot ({\bf \bar{A}}_-^*(t)\times \hat{n})) \right\} \nonumber \\
  & = & \frac{9}{16\pi} \left\{ |{\bf A_+}\times \hat{n}|^2 |\bar{f}_+(t)|^2 +  |{\bf A_-}\times \hat{n}|^2 |\bar{f}_-(t)|^2 \right. \nonumber \\
  &  & + \left.2 Re(({\bf A_+}\times \hat{n}) \cdot ({\bf A_-^*}\times \hat{n})\cdot \bar{f}_+(t)\cdot \bar{f}_-^*(t)\right\}
\label{eqn:fullProb2}
\end{eqnarray}
% ++++++++++++++++++++++++++++++++++++++++++++++++++++++++++++++++++++++
where  (for ${\bar B}$) the diagonal term in Eq.~(\ref{eqn:fullProb2}) is
% ++++++++++++++++++++++++++++++++++++++++++++++++++++++++++++++++++++++
\begin{eqnarray}
|\bar{f_\pm}(t)|^2 =\frac{1}{2} \frac{ (1 \pm \cos{2\beta_s}) e^{-\Gamma_Lt} + (1\mp \cos{2\beta_s})e^{-\Gamma_Ht}\pm   {2\sin{2\beta_s} e^{-\Gamma t}\sin{\Delta m t}} }{\tau_L(1 \pm \cos{2\beta_s}) + \tau_H(1 \mp \cos{2\beta_s})}\,,\nonumber \\
\label{eqn:fullProb3A}
\end{eqnarray}
while  (for $B$) the diagonal term in Eq.~(\ref{eqn:fullProb1}) is
\begin{eqnarray}
 |f_\pm(t)|^2 =\frac{1}{2}\frac{ (1 \pm \cos{2\beta_s}) e^{-\Gamma_Lt} +(1\mp \cos{2\beta_s}) e^{-\Gamma_Ht} \mp   {2\sin{2\beta_s}e^{-\Gamma t} \sin{\Delta m t}} }{\tau_L(1 \pm \cos{2\beta_s}) + \tau_H(1 \mp \cos{2\beta_s})} \nonumber \\
\label{eqn:fullProb3B}
\end{eqnarray}
% ++++++++++++++++++++++++++++++++++++++++++++++++++++++++++++++++++++++
and (for $\bar{B}$) the cross-term, or interference term in Eq.~(\ref{eqn:fullProb2}) is
% ++++++++++++++++++++++++++++++++++++++++++++++++++++++++++++++++++++++
\begin{eqnarray}
\bar{f_+}(t) \bar{f_-}^*(t) =   \frac{ -e^{-\Gamma t}\cos{\Delta m t} -i\cos{2\beta_s}e^{-\Gamma t}\sin{\Delta m t} +i\sin{2\beta_s} (e^{-\Gamma_L t} - e^{-\Gamma_H t})/2}{\sqrt{\left[ (\tau_L-\tau_H)\sin{2\beta_s}\right]^2+4\tau_L\tau_H}}\,, \nonumber \\
\label{eqn:fullProb4A}
\end{eqnarray}
while (for $B$) the interference term in Eq.~(\ref{eqn:fullProb1}) is
\begin{eqnarray}
f_+(t) f_-^*(t) = \frac{ e^{-\Gamma t}\cos{\Delta m t} +i\cos{2\beta_s}e^{-\Gamma t}\sin{\Delta m t} +i\sin{2\beta_s} (e^{-\Gamma_L t} - e^{-\Gamma_H t})/2}{\sqrt{\left[ (\tau_L-\tau_H)\sin{2\beta_s}\right]^2+4\tau_L\tau_H}}\,. \nonumber \\
\label{eqn:fullProb4B}
\end{eqnarray}
% ++++++++++++++++++++++++++++++++++++++++++++++++++++++++++++++++++++++
This accomplishes the desired separation.  In the fitting function, to
accommodate the proper time resolution, one has only to replace all time-dependent functions with
their smeared equivalents. 

\section {Sensitivity to $\Delta m$}

It can be noticed that the time development of the interference term, expressions~\ref{eqn:fullProb4A} and \ref{eqn:fullProb4B}, 
contain undiluted mixing asymmetries \emph{even in the case of no CP violation}, i.e., when $\beta_s=0$.  
Let us try to better understand the mechanism by which the flavor of the $B^0_s$ meson is tagged at \emph{decay}
time, by first rewriting Eq.~(\ref{eqn:startingPoint}) using the $B_s^H$ and $B_S^L$
states in the expansion rather than the $B^0_s$ and ${\bar B}^0_s$ states:
\begin{eqnarray}
          \langle& \mu^+\mu^-&K^+K^-|H|B^0_{s,phys}(t)\rangle  =  \nonumber \\
&         & \sum_{i} {\braket{\mu^+\mu^-K^+K^-|H|{\cal P}_i}  \braket{{\cal P}_i|H|B_s^H}        \braket{B_s^H|B^0_{s,phys}(t)}       }  \nonumber \\
&  +      &\sum_{i}{\braket{\mu^+\mu^-K^+K^-|H|{\cal P}_i}  \braket{{\cal P}_i|H|B_s^L}       \braket{B_s^L|B^0_{s,phys}(t)}}. 
\label{eqn:bigsum}
\end{eqnarray}
Now, we take the limit of zero $CP$ violation in the $B^0_s$ system, such that $\braket{{\cal P}_{||}|H|B_s^H} = \braket{{\cal P}_{0}|H|B_s^H} = \braket{{\cal P}_{\perp}|H|B_s^L} = 0$, and only three of the six terms in Eq.~(\ref{eqn:bigsum}) remain:
\begin{eqnarray}
\langle&\mu^+\mu^-&K^+K^-|H|B^0_{s,phys}(t)\rangle  = \nonumber \\
& & {\braket{\mu^+\mu^-K^+K^-|H|{\cal P}_{\perp}}  \braket{{\cal P}_{\perp}|H|B_s^H}        \braket{B_s^H|B^0_{s,phys}(t)}           }  \nonumber \\
                                       & + & {\braket{\mu^+\mu^-K^+K^-|H|{\cal P}_{0}}      \braket{{\cal P}_{0}|H|B_s^L}        \braket{B_s^L|B^0_{s,phys}(t)}           }  \nonumber  \\ 
                                       & + & {\braket{\mu^+\mu^-K^+K^-|H|{\cal P}_{||}}     \braket{{\cal P}_{||}|H|B_s^L}        \braket{B_s^L|B^0_{s,phys}(t)}           } .
\end{eqnarray}
When the expression is squared, the interference terms are the cross terms involving both the product of a $CP$-even and a $CP$-odd amplitudes. The time dependence of these terms is contained in 
the factor:
\begin{eqnarray}
 \langle &B_s^H&|B_{s,phys}^0(t)\rangle\langle B_s^L|B_{s,phys}^0(t)\rangle  =\nonumber \\
& &   \frac{1}{4} \left [ \left( \braket{B_s^H|B_{s,phys}^0(t)}+ \braket{B_s^L|B_{s,phys}^0(t)}\right)^2 \right. \nonumber \\
&  & -\left. \left( \braket{B_s^H|B_{s,phys}^0(t)}- \braket{B_s^L|B_{s,phys}^0(t)}\right)^2 \right]   \nonumber \\
& = &                       \frac{1}{2}\left [\left( \frac{\bra{B_s^H}+\bra{B_s^L}}{\sqrt{2}}\ket{B_{s,phys}^0(t)}\right)^2 -  \left(\frac{\bra{B_s^H}-\bra{B_s^L}}{\sqrt{2}}\ket{B_{s,phys}^0(t)}\right )^2 \right]  \nonumber \\
& = &                       \frac{1}{2}\left [\braket{B^0_s|B_{s,phys}^0(t)}^2 - \braket{\bar{B}^0_s|B_{s,phys}^0(t)}^2 \right ]. \nonumber \\
\end{eqnarray}
This factor takes the value +1/2 when the meson is pure $B^0_s$, and -1/2 when the
meson is pure ${\bar B}^0_s$, and in general oscillates between these two values.  Thus the
interference term effectively tags the flavor of the $B^0_s$ at decay.  This provides a way to 
observe $B^0_s \rightarrow \bar {B}^0_s$ flavor oscillations using a sample of  flavor-tagged 
$B^0_s\rightarrow J/\psi \phi$ decays which can be collected with a simple dimuon trigger.  
This may open a particularly interesting avenue for the LHC experiments to 
observe $B^0_s$ mixing using a $J/\psi$ trigger.

\section{Incorporating Direct $CP$ Violation}

An asymmetry either in the decay rate ($|\bar{A_i}/A_i|\ne$ 1) or in
the mixing ($|q/p|\ne$ 1) such that $|\lambda|\ne 1$ is direct $CP$
violation.  In the case of direct $CP$ violation $\lambda$ does not lie on the
unit circle in the complex plane, and we need two parameters to
describe it which we will take to be ${\cal C}\equiv Re(\lambda)$ and
${\cal S}\equiv Im(\lambda)$.  Experimentally, even if one sets out
to extract $\beta_s$ assuming the constraint $|\lambda|=1$, it is
nonetheless of interest to test that constraint, since sensitivity to
${\cal C}$ and ${\cal S}$ arise from very different features of the
detector.  In that case we must revisit not only the functional form
of the differential decay rates, but also the normalization. The amplitudes
in Eq.~(\ref{eqn:finalAmp}) must now be written as:

\begin{eqnarray}
{\cal A}_i &=& {\cal N}_{\pm}{e^{-\Gamma t /2}} \left[E_+(t) \pm \lambda E_-(t) \right] a_i\,,   \nonumber \\
{\bar {\cal A}_i} &=& {\cal N}_{\pm}{e^{-\Gamma t /2}}\left[\pm E_+(t) + E_-(t)/\lambda \right] a_i\,,  
\label{eqn:dcpvAmp}
\end{eqnarray}
where 
\begin{eqnarray*}
{\cal N}_{\pm}  &=&  \left \{\frac {1}{4|\lambda|^2} \left [\left [ (\tau_H + \tau_L) (1+|\lambda|^2)^2 \pm 2 {\cal C} \cdot (\tau_L - \tau_H) (1+|\lambda|^2) \right ]  \right. \right. \nonumber \\
& & \left. \left.   +  \frac{\tau}{1+\Delta m^2 \tau^2}  \cdot  \left [   
\pm 4{\cal S}\cdot \left ( 1-|\lambda|^2 \right )  \Delta m\tau - 2\left(1-|\lambda|^2 \right )^2
\right ]            \right ] \right \}^{-\frac{1}{2}} \,.
\end{eqnarray*}
These amplitudes can readily be seen to reduce to those of Eq.~(\ref{eqn:finalAmp}) in the limit of $ {\cal C}^2 + {\cal S}^2 \equiv |\lambda|^2 \rightarrow 1$. The normalization of detector efficiency, Eq.~(\ref{eqn:normFactor}), becomes:
\begin{eqnarray*}
  N  = & & \frac{3}{8\sqrt{ \pi}}   \left[   \frac{4c_{00}^0}{3}(|a_0|^2 + |a_{\parallel}|^2 + |a_{\perp}|^2) \right. \nonumber \\
& + &     \left.  \frac{4c_{00}^2}{15}(2|a_0|^2 - |a_{\parallel}|^2 - |a_{\perp}|^2)   \right]  \\
 &   + &  \frac{3}{8\sqrt{5\pi}}   \left[   \frac{2c_{20}^0}{3}( |a_0|^2 + |a_{\parallel}|^2 - 2|a_{\perp}|^2) \right. \nonumber \\
  &  + &   \left. \frac{4c_{20}^2}{15}(|a_0|^2 - \frac{1}{2}|a_{\parallel}|^2 + |a_{\perp}|^2)   \right]  \\
 & - &  \frac{9}{16\sqrt{15\pi}}                   {\cal N}_+{\cal N}_-{\cal S}\cdot (\tau_L-\tau_H) \nonumber \\
 & &\times  \left[(a_{\parallel}^*a_{\perp}+a_{\parallel}a_{\perp}^*)(\frac{4}{3} c_{2-1}^0 - \frac{4}{15} c_{2-1}^2)    \right] \\
 &    + &  \frac{9}{16} \frac{\sqrt{2}}{\sqrt{15\pi}}   {\cal N}_+{\cal N}_-{\cal S}\cdot(\tau_L-\tau_H) \nonumber \\
& &  \times \left[(a_{0}^*a_{\perp}+a_{0}a_{\perp}^* ) (\frac{\pi c_{21}^{1}}{8} - \frac{\pi c_{21}^3}{32} + ... ) \right]\\ 
  &   + & \frac{9}{8\sqrt{15\pi}}   \left[   \frac{2c_{22}^0}{3}(-|a_0|^2 +|a_{\parallel}|^2) - \frac{4c_{22}^2}{15}(|a_0|^2 + \frac{1}{2}|a_{\parallel}|^2)   \right]  \\
  &   + & \frac{9}{16} \frac{\sqrt{2}}{\sqrt{15\pi}}  \left[(a_{0}^*a_{\parallel}+a_{0}a_{\parallel}^*) (\frac{\pi c_{2-2}^{1}}{8} - \frac{\pi c_{2-2}^3}{32} + ... ) \right].
\end{eqnarray*}
Finally, the explicit time development, Eqs.~(\ref{eqn:fullProb3A}), (\ref{eqn:fullProb3B}), (\ref{eqn:fullProb4A}) and (\ref{eqn:fullProb4B}), must be replaced with the more general forms:
\begin{eqnarray*}
 |\bar{f_\pm}(t)|^2 & = & \frac{{\cal N}_{\pm}^2}{4 |\lambda|^2}\left [((1+|\lambda|^2) \pm 2{\cal C}) e^{-\Gamma_Lt} + ((1+|\lambda|^2) \mp 2{\cal C})e^{-\Gamma_Ht} \right . \nonumber \\
& & \left. +    \left (\pm 4{\cal S} \sin{\Delta m t} -2 (1-|\lambda|^2) \cos{\Delta m t} \right )  e^{-\Gamma t}   \right ], \nonumber \\
%\end{eqnarray*}
%\begin{eqnarray*}
  |f_\pm(t)|^2 &=& \frac{{\cal N}_{\pm}^2}{4}\left [ ((1+|\lambda|^2) \pm 2{\cal C}) e^{-\Gamma_Lt} + ((1+|\lambda|^2) \mp 2{\cal C})e^{-\Gamma_Ht} \right. \nonumber \\
& & \left. - \left (\pm 4{\cal S} \sin{\Delta m t} -2 (1-|\lambda|^2) \cos{\Delta m t} \right )  e^{-\Gamma t} \right ], \nonumber \\
%\end{eqnarray*}
%\begin{eqnarray*}
  \bar{f_+}(t) \bar{f_-}^*(t) & = &  \frac{{\cal N}_+{\cal N}_-}{4|\lambda|^2} \left [-e^{-\Gamma t}\left(  2(1+|\lambda|^2)\cos{\Delta m t} + 4i{\cal C} \sin {\Delta m t} \right) \right. \nonumber \\
& & \left. +e^{-\Gamma_L t}\left((1-|\lambda|^2) + 2i{\cal S}\right) + e^{-\Gamma_H t}\left((1-|\lambda|^2) - 2i{\cal S}\right)  \right ], \nonumber \\
%\end{eqnarray*}
%and
%\begin{eqnarray*} 
  f_+(t)  f_-^*(t) & = &  \frac{{\cal N}_+{\cal N}_-}{4} \left [e^{-\Gamma t}\left(  2(1+|\lambda|^2)\cos{\Delta m t} + 4i{\cal C} \sin {\Delta m t} \right) \right. \nonumber \\
& & \left. +e^{-\Gamma_L t}\left((1-|\lambda|^2) + 2i{\cal S}\right) + e^{-\Gamma_H t}\left((1-|\lambda|^2) - 2i{\cal S}\right)  \right ], \nonumber \\
\end{eqnarray*}

% ++++++++++++++++++++++++++++++++++++++++++++++++++++++++++++++++++++++
\noindent which can be seen to reduce to expression~\ref{eqn:fullProb3A}, \ref{eqn:fullProb3B} and \ref{eqn:fullProb4A}, \ref{eqn:fullProb4B}
as $|\lambda|^2 \rightarrow 1$.

\section{Incorporating a Contribution from $B^0_s \rightarrow J/\psi K^+K^-$ (Kaons in an $S$-Wave State)}

It has been suggested~\cite{ref:Sheldon} that a contribution 
from $S$-wave $K^+K^-$ under the $\phi$ peak in $B^0_s\rightarrow J/\psi\phi$ decay may contribute up to 5-10\% of the total rate. A normalized probability density for the decay $B^0_s \rightarrow J/\psi K^+K^-$ (kaons in an
$S$-wave state) can
be worked out by considering the polarization vector of the $J/\psi$ in the decay and proceeding as
in \cite{ref:dighe}. The resulting expressions
\begin{eqnarray} 
Q_{B}(\theta, \varphi, \psi, t) &=& \frac{3}{16\pi} |{\bf B}(t)\times \hat{n}|^2\,,  \nonumber \\
Q_{\bar{B}}(\theta, \varphi, \psi, t) &=& \frac{3}{16\pi} |{\bf {\bar B}(t)}\times \hat{n}|^2
\label{eqn:PureS}
\end{eqnarray}
do not depend at all on the angle $\psi$ (which is the helicity angle in the $\phi$ decay).
In the previous expression
\begin{eqnarray}
{\bf B}(t)& = &\left({\mathcal B}(t),0,0\right) \,, \nonumber \\
{\bf{\bar {B}}}(t) & =&\left({\bar {\mathcal B}}(t),0,0\right)
\label{eqn:fixedAngleS}
\end{eqnarray}
where the time-dependent amplitudes,
\begin{eqnarray}
{\cal B}(t) &=& \frac{e^{-\Gamma t /2}}{\sqrt{\tau_H + \tau_L - \cos{2\beta_s}\left(\tau_L-\tau_H\right)}} \left[
E_+(t) - e^{2i\beta_s} E_-(t)
\right],   \nonumber \\
{\bar {\cal B}}(t) &=& \frac{e^{-\Gamma t /2}}{\sqrt{\tau_H + \tau_L - \cos{2\beta_s}\left(\tau_L-\tau_H\right)}} \left[
-E_+(t) + e^{-2i\beta_s} E_-(t)
\right]  \nonumber \\
\label{eqn:finalAmpS}
\end{eqnarray}
reflect the $CP$-odd nature of the $J/\psi KK$ final state. 

When both $P$-wave and $S$-wave are present, the amplitudes must be summed and then squared.  The $P$ wave has a resonant
structure due to the $\phi$-propagator, while the $S$-wave amplitude is flat (but can have any phase with respect the 
$P$-wave).  Suppose that in our experiment we accept events for which the reconstructed mass $m(K^+K^-) \equiv \mu$ lies within a window $ \mu_{lo} < \mu < \mu_{hi}$. 
The normalized probability in this case is
\begin{eqnarray} 
\rho_{B}(\theta, \varphi, \psi, t, \mu) &=& \frac{9}{16\pi} \left |\left      [ \sqrt{1-F_s}g(\mu){\bf A}(t) + e^{i\delta_s}\sqrt{F_s}\frac {h(\mu)}{\sqrt{3}}{\bf B}(t)\right ]\times \hat{n} \right |^2  \,, \nonumber \\
\rho_{\bar{B}}(\theta, \varphi, \psi, t, \mu) &=& \frac{9}{16\pi} \left |\left[ \sqrt{1-F_s}g(\mu) {\bf {\bar A}} (t) + e^{i\delta_s}\sqrt{F_s}\frac {h(\mu)}{\sqrt{3}}{\bf {\bar B}(t)} \right ]\times \hat{n}\right |^2 \,,\nonumber \\ 
\label{eqn:PAndS}
\end{eqnarray}
where we use a nonrelativistic Breit-Wigner to model the $\phi$ resonance\footnote{We shall have more
to say about that, later.} 
\begin{equation}
  g(\mu) = \sqrt{\frac {\Gamma_{\phi}/2}{\Delta \omega}} \cdot \frac{1}{\mu - \mu_{\phi} + i \Gamma_{\phi}/2}
\label{eqn:gDef}
\end{equation}
a flat model for the $S$-wave mass distribution
\begin{equation}
h(\mu) = \frac {1}{\sqrt{\Delta \mu}}
\label{eqn:hDef}
\end{equation}
and define
\begin{equation}
\omega_{hi} = \tan^{-1}{\frac {2(\mu_{hi}- \mu_{\phi})}{\Gamma_{\phi}}}
\quad
\omega_{lo} = \tan^{-1}{\frac {2(\mu_{lo}- \mu_{\phi})}{\Gamma_{\phi}}}
\end{equation}
and 
\begin{equation}
\Delta \mu = \mu_{hi} - \mu_{lo}
\qquad
\Delta \omega = \omega_{hi}-\omega_{lo} \,.
\end{equation}
In these equations, $F_s$ is the $S$-wave fraction; $\mu_{\phi}$ is the $\phi$ 
mass (1019 MeV/c$^2$); $\Gamma_{\phi}$ is the $\phi$ width (4.26 MeV/c$^2$),
and $\delta_{s}$ is the phase of the $S$-wave component relative to 
the $P$-wave component.

In the presence of an $S$-wave contribution, the normalization of Eq.~(\ref{eqn:normFactor}) must be generalized; in order to do this we first define the quantities 
\begin{equation}
{\cal F}(\mu) \equiv 
\sqrt{\frac{F_s (1-F_s) \Gamma_{\phi}}{2 \Delta \mu \Delta \omega}} \cdot
\frac{e^{-i\delta_s}}{\mu-\mu_{\phi}+i\Gamma_{\phi}/2}
\end{equation}
and
\begin{eqnarray}
& & {\cal I}_\mu \equiv \int {\cal F}(\mu) d\mu = \nonumber \\
& & \sqrt{\frac{F_s (1-F_s)\Gamma_{\phi}}{2 \Delta \mu \Delta \omega}} \cdot
e^{-i\delta_s} \cdot \log { \frac {\mu_{hi}-\mu_{\phi}+i\Gamma_{\phi}/2}{\mu_{lo}-\mu_{\phi}+i\Gamma_{\phi}/2} } \,. \nonumber \\
\label{eqn:Imu}
\end{eqnarray}
Then the normalizing factor appropriate for Eq.~(\ref{eqn:PAndS}) is 
\begin {equation}
{\cal N }  =  (1-F_s) \cdot N +  2 Re \left [ {\cal I}_\mu \cdot N^\prime \right ] +  F_s\cdot N^{\prime\prime} 
\end{equation}
where $N$ is given in Eq.~(\ref{eqn:normFactor}), and
\begin{eqnarray}
N^{\prime}  = & & \sqrt{3}*a_0^*(\frac{1}{6\sqrt{\pi}}c^1_{00} + 
\frac{1}{12\sqrt{5\pi}}c^1_{20} -
\frac{1}{4\sqrt{15\pi}}c^1_{22}) \nonumber \\
& + & \frac{3}{16} \sqrt{\frac{2}{5\pi}}a_{\parallel}^*(\frac{\pi}{2} c^0_{2-2} - \frac{\pi}{8} c^2_{2-2} + ...) \nonumber \\
& + & \frac{3}{16} \sqrt{\frac{2}{5\pi}}a_{\perp}^{*}  \frac{\sin{2\beta_s} (\tau_L-\tau_H)}{   \sqrt{ ((\tau_L  -  \tau_H)\sin{2\beta_s})^2   +{4\tau_L\tau_H} }} (\frac{\pi}{2} c^0_{2 1} - \frac{\pi}{8} c^2_{2 1} + ...) \nonumber \\
%N^{\prime} & = & \frac{1}{2\sqrt{\pi}}\cdot a_0\cdot (\frac{2}{3}c^1_{00} + \frac{1}{\sqrt{15}}c^1_{20} + \frac{1}{\sqrt{45}}c^1_{22}) \nonumber \\ 
%& - &
%\frac{3}{8}\sqrt{\frac{2}{5\pi}}\cdot \left ( a_{\perp} \frac{\sin{2\beta_s} (\tau_L-\tau_H)}{   \sqrt{ ((\tau_L  -  \tau_H)\sin{2\beta_s})^2   +{4\tau_L\tau_H} }} - a_\parallel \right ) \nonumber\\
%& \times &  
%\left( 
%8 (c^0_{21}+c^0_{2,-2}) - \frac{1}{15} (c^2_{21}+c^2_{2,-2}) + ...  
%\right) 
\label{eqn:Farrukh}
\end{eqnarray}
and
\begin {equation}
N^{\prime\prime} = \frac{1}{2\sqrt{\pi}}c^0_{00} + 
\frac{1}{4\sqrt{5\pi}}c^0_{20} -
\frac{3}{4\sqrt{15\pi}}c^0_{22}\,.
\end{equation}
The numerical factors $+\pi/2$ and $-\pi/8$ appearing together with $c^k_{2,1}$ and $c^k_{2,-2}$ in the infinite series 
are the integrals
\begin {equation}
\int P_k(\cos{\psi}) \sin{\psi} d(\cos{\psi})\,.
\end{equation}

We now work out the explicit time and mass dependence of the differential rates.  We will use Eq.~(\ref{eqn:fullProb1})
together with the analogous equation for the pure $S$-wave differential rate:

\begin{eqnarray}
Q_{B}(\theta, \psi, \varphi, t)  & = & \frac{3}{16\pi} |{\bf B(t)}\times \hat{n}|^2 \nonumber \\
  & = & \frac{3}{16\pi} |{\bf B}\times \hat{n}|^2 |f_-(t)|^2
\label{eqn:swaveProb1}
\end{eqnarray}
% ++++++++++++++++++++++++++++++++++++++++++++++++++++++++++++++++++++++
and
% ++++++++++++++++++++++++++++++++++++++++++++++++++++++++++++++++++++++
\begin{eqnarray}
Q_{\bar{B}}(\theta, \psi, \varphi, t)  & = & \frac{3}{16\pi} |{\bf \bar{B}}(t)\times \hat{n}|^2 \nonumber \\
  & = & \frac{3}{16\pi} |{\bf B}\times \hat{n}|^2 |\bar{f}_-(t)|^2\,.
\label{eqn:swaveProb2}
\end{eqnarray}
where the vector ${\bf B} = \hat{x} = (1,0,0)$.  The full probability densities, which can be used in a time-, angle-, and $\phi$ mass-dependent fit, are obtained by expanding Eq.~(\ref{eqn:PAndS}). We get
\begin{eqnarray}
&   \rho_{B}&(\theta, \psi, \varphi, t,\mu ) = \nonumber \\
& &  (1-F_s)\frac {\Gamma_{\phi}/2}{\Delta \omega} \cdot \frac{1}{(\mu - \mu_{\phi})^2 + \Gamma_{\phi}^2/4}\cdot P_{B}(\theta, \psi, \varphi, t) \nonumber \\
& & +  F_s\frac{1}{\Delta \mu} Q_{B}(\theta, \psi, \varphi, t)  \nonumber \\
                                    & & +   2\frac {\sqrt{27}}{16\pi} Re \left [{\cal F}(\mu) \left (({\bf A_-}\times \hat{n})\cdot ({\bf B}\times \hat{n}) \cdot |f_-(t)|^2 \right. \right. \nonumber \\
& & \left. \left. +  ({\bf A_+}\times \hat{n}) \cdot ({\bf B}\times \hat{n})\cdot f_+(t)\cdot f_-^*(t) \right )\right ]\nonumber \\
\label{eqn:comboProb1}
\end{eqnarray}
and
\begin{eqnarray}
&\rho_{\bar{B}}&(\theta, \psi, \varphi, t,\mu )   = \nonumber \\
& &   (1-F_s)\frac {\Gamma_{\phi}/2}{\Delta \omega} \cdot \frac{1}{(\mu - \mu_{\phi})^2 +  \frac{1}{1+F_S}\Gamma_{\phi}^2/4}\cdot P_{\bar {B}}(\theta, \psi, \varphi, t) \nonumber \\
& & +  F_s\frac{1}{\Delta \mu} Q_{\bar {B}}(\theta, \psi, \varphi, t)  \nonumber \\
&  & +   2 \frac {\sqrt{27}}{16\pi} Re \left [{\cal F}(\mu) \left (  ({\bf A_-}\times \hat{n})\cdot({\bf B}\times \hat{n}) \cdot |{\bar f}_-(t)|^2 \right. \right. \nonumber \\
& & \left. \left. +  ({\bf A_+}\times \hat{n}) \cdot ({\bf B}\times \hat{n})\cdot {\bar f}_+(t)\cdot {\bar f}_-^*(t) \right )\right ] \,.\nonumber \\
\label{eqn:comboProb2}
\end{eqnarray}
\clearpage
In case one does not want to observe the $\phi$-mass variable $\mu$, one can integrate it out.  Then one obtains

\begin{eqnarray}
&\rho_{B}&(\theta, \psi, \varphi, t)   =  \nonumber \\
& &  (1-F_s)\cdot P_{B}(\theta, \psi, \varphi, t) +  F_s Q_{B}(\theta, \psi, \varphi, t)  \nonumber \\
                                    & & +  2 \frac {\sqrt{27}}{16\pi} Re \left [{\cal I}_\mu \left (  ({\bf A_-}\times \hat{n})\cdot ({\bf B}\times \hat{n}) \cdot |f_-(t)|^2 \right. \right. \nonumber \\
& & \left. \left. +  ({\bf A_+}\times \hat{n}) \cdot ({\bf B}\times \hat{n})\cdot f_+(t)\cdot f_-^*(t) \right )\right ], \nonumber \\
\label{eqn:comboProbInt1}
\end{eqnarray}

\begin{eqnarray}
&\rho_{\bar{B}}&(\theta, \psi, \varphi, t)   =  \nonumber \\
& &  (1-F_S)\cdot P_{\bar {B}}(\theta, \psi, \varphi, t) +  F_s Q_{\bar {B}}(\theta, \psi, \varphi, t)  \nonumber \\
                                    & & +  2 \frac {\sqrt{27}}{16\pi} Re \left [{\cal I}_\mu \left ( ({\bf A_-}\times \hat{n})\cdot ({\bf B}\times \hat{n}) \cdot |{\bar f}_-(t)|^2 \right. \right. \nonumber \\
& & \left. \left. +  ({\bf A_+}\times \hat{n}) \cdot ({\bf B}\times \hat{n}) \cdot {\bar f}_+(t)\cdot {\bar f}_-^*(t) \right )\right ]. \nonumber \\
\label{eqn:comboProbInt2}
\end{eqnarray}

\section{Symmetries}
\label{sec:symmetries}
In this section we examine the symmetries of our differential rate formulae, starting from the simplest
case, $K^+K^-$ in a $P$-wave, Eq.~(\ref{eqn:finalAngle}), but considering also the case where both $P$ and $S$ waves are included, Eq.~(\ref{eqn:PAndS}).  In the case of pure $P$-wave, one can readily spot that the probability densities in
Eq.~(\ref{eqn:finalAngle}) are invariant to the following transformations:
\begin{itemize}
\item {A simultaneous rotation of the vectors  ${\bf A}(t)$ and ${\hat n}$}
\item {An inversion of the vector ${\bf A}(t)$}
\item {Complex-conjugation of the vector ${\bf A}(t)$}
\end{itemize}

The symmetry to simultaneous rotation of the vectors ${\bf A(t)}$ and
${\hat n}$ corresponds to the well-known freedom to choose a
convenient basis in which to work.  An example of an alternative basis
is the \emph{helicity basis}, which derives from the \emph{transversity
basis} by a cyclic permutation of the coordinate axis: ${\hat x}_T =
{\hat z}_H$, etc.  One can take the angles in Eq.~(\ref{eqn:nhat}) to be
the polar and azimuthal angles in the helicity basis, but then one
must transform ${\bf A(t)}$ accordingly, i.e, by permuting the
elements of ${\bf A(t)}$ in the defining equation,
Eq.~(\ref{eqn:fixedAngle}).  Then, Eq.~(\ref{eqn:finalAngle}) remains
valid in the helicity basis.  This rotational invariance implies that
the choice of basis is irrelevant to the final result since the  
likelihood is invariant to the choice (though we do not rule out
the possibility that the quality of the efficiency expansion,  Eq.~(\ref{eqn:Pprimeprime}), 
may depend on the choice of basis, as pointed out in \cite{ref:GronauRosner}).
 
A more interesting symmetry is the symmetry that results from
transforming ${\bf A}(t)$ to its complex conjugate.  If we take, by 
convention, $a_0$ to be real and let $\delta_\parallel = \arg(a_\parallel)$,
and $\delta_\perp = \arg(a_\perp)$, then as we will demonstrate below,
this conjugation transformation is equivalent to the simultaneous transformation:
%\begin{itemize}
%\item {$\beta_s \rightarrow \pi/2 - \beta_s$}
%\item {$\Delta\Gamma \rightarrow-\Delta\Gamma$}
%\item {$\delta_{\perp} = \pi - \delta_{\perp}$}
%\item {$\delta_{\parallel} = 2\pi - \delta_{\parallel}$}
%\end{itemize}  
\begin{eqnarray}
\beta_s &\rightarrow& \pi/2 - \beta_s \nonumber \\
\Delta\Gamma &\rightarrow &-\Delta\Gamma \nonumber \\
\delta_{\perp} &\rightarrow& \pi - \delta_{\perp} \nonumber \\
\delta_{\parallel} &\rightarrow& 2\pi - \delta_{\parallel} \,. 
\end{eqnarray}  
That is to say that the simultaneous transformation of these four
variables is a symmetry of the likelihood \emph{because} it transforms
${\bf A(t)}$ into its complex conjugate.  Since for pure $P$ wave state the
combined transformation is a well-known symmetry, this observation may
appear as a curiosity; however when both $P$ and $S$ wave states are
included, we shall see that complex conjugation teaches us how to
properly extend the symmetry. First, we show how the combined
transformation accomplishes the claimed complex conjugation.
\begin {enumerate}
\item {Note from Eq.~(\ref{eqn:functionDef}) that the combined transformation
transforms $E_{\pm}(t) \rightarrow \pm  E_{\pm}^*(t)$. }
\item {Note also that the combined transformation transforms $e^{-2i\beta_s}
\rightarrow -e^{+2i\beta_s}$ and $e^{+2i\beta_s}
\rightarrow -e^{-2i\beta_s}$}
\item {Therefore, in Eq.~(\ref{eqn:finalAmp}), the terms in square brackets
are transformed into their complex conjugates}. 
\item {Note that both $\cos{2\beta_s}$ and $\tau_L-\tau_H$  
change sign under the transformation, so also the piece of Eq.~(\ref{eqn:finalAmp}) 
in the denominator, under the square root sign, is invariant under the 
combined transformation; since that piece is real we can say that it is
anyway equal to its complex conjugate.}
\item {The real quantity $a_0$ does not change under the combined
transformation, but since it is real, it is anyway equal to $a_0^*$.}
\item {The combined transformation transforms $a_\parallel \rightarrow a_\parallel^*$.}
\item {The combined transformation transforms $i a_\perp \rightarrow -i a_\perp^*$.}
\item{Then looking at Eq.~(\ref{eqn:fixedAngle}), one sees finally that the net
effect of the combined transformation has been the complex conjugation of 
the vector ${\bf A(t)}$. }
\end {enumerate}

Returning now to the full  likelihood including
both $P$ and $S$ wave states,
Eq.~(\ref{eqn:PAndS}), we can see that, here again, complex
conjugation of the term 
\begin{equation}
\sqrt{1-F_s}g(\mu){\bf A}(t) +e^{i\delta_s}\sqrt{F_s}\frac {h(\mu)}{\sqrt{3}}{\bf B}(t)
\label{eqn:someTerms}
\end{equation}
leaves the probability density invariant (in a parameter space now enlarged 
to include $\mu_\phi$ and $\Gamma_{\phi}$); however now, complex conjugation of the term
$g(\mu)$, Eq.~(\ref{eqn:gDef}), implies that the transformation 
$\Gamma_{\phi} \rightarrow -\Gamma_{\phi}$ should also be carried out, in addition 
to the transformation of $\beta_s$, $\Delta\Gamma$, $\delta_{\parallel}$, 
and $\delta_{\perp}$.  Since negative values of $\Gamma_{\phi}$ are physically
meaningless, this transformation is not an admissible symmetry. 

However we can find a symmetry transformation that carries one set of physically 
meaningful parameters into another.  Such a symmetry is the transformation
of the terms in  Eq.~(\ref{eqn:someTerms}) into their \emph{negative complex 
conjugate}.  This transformation is equivalent to the combined transformation
already described, in addition to:
%\begin{itemize}
%\item {$\delta_s \rightarrow \pi - \delta_s$}
%\item {$(\mu-\mu_\phi) \rightarrow -(\mu - \mu_\phi)$.}
%\end{itemize}
\begin{eqnarray}
\delta_s &\rightarrow& \pi - \delta_s \nonumber \\
(\mu-\mu_\phi) &\rightarrow& -(\mu - \mu_\phi) \,. 
\end{eqnarray}
The latter transformation carries us from a point on one side of the $\phi$
mass peak to another point located symmetrically on the other side.  This
symmetry is useful when considering likelihood functions in which
the dependence on $\mu$ is integrated out.  If we
integrate symmetrically about the $\phi$ mass peak, we can consider the contribution to the integral
coming from a slice in $\phi$ mass on one hand and the a symmetrically-located
slice in $\phi$ mass on the other hand.  While the contribution of either slice
is not invariant to the transformation above, the contribution of both
slices certainly is, and the combined transformation:
%\begin{itemize}
%\item {$\beta_s \rightarrow \pi/2 - \beta_s$}
%\item {$\Delta\Gamma \rightarrow-\Delta\Gamma$}
%\item {$\delta_{\perp} = \pi - \delta_{\perp}$}
%\item {$\delta_{\parallel} = 2\pi - \delta_{\parallel}$}
%\item {$\delta_s \rightarrow \pi - \delta_s$}
%\end{itemize}  
\begin{eqnarray}
\beta_s &\rightarrow& \pi/2 - \beta_s \nonumber \\
\Delta\Gamma &\rightarrow& -\Delta\Gamma \nonumber \\
\delta_{\perp} &\rightarrow& \pi - \delta_{\perp} \nonumber \\
\delta_{\parallel} &\rightarrow& 2\pi - \delta_{\parallel} \nonumber \\
\delta_s &\rightarrow& \pi - \delta_s 
\label{eqn:symmetries}
\end{eqnarray}  
is again a symmetry of the integrated likelihood.  We note, however, that this
symmetry requires the symmetry of the nonrelativistic $\phi$-propagator, Eq.~(\ref{eqn:gDef}), and applies to {\it the likelihood integrated over a finite symmetric interval of integration}.  

Symmetries of the likelihood function  for $B^0_s \rightarrow J/\psi \phi$, 
in the presence of $S$-wave contribution for a fixed value of  $\mu=m(K^+K^-)$
were discussed in a recent publication~\cite{ref:LHCb}. These formula can
also used to fit for data falling within a narrow window in $\mu$.  Under those assumptions we can drop the $\phi$ propagator from 
the expression in Eq.~(\ref{eqn:someTerms}), absorb the Breit-Wigner terms into
the amplitudes ${\bf A}(t)$, and write our model for the rates as
\begin{equation}
\sqrt{1-F_s}{\bf A}(t) +e^{i\delta_s}\sqrt{\frac{F_s}{3}}{\bf B}(t) \,.
\label{eqn:otherTerms}
\end{equation}
Then one can see that the transformation in which $\delta_s \rightarrow - \delta_s$  replaces $\delta_s \rightarrow \pi - \delta_s$ in Eq.~\ref{eqn:symmetries} accomplishes a complex conjugation of the terms in Eq.~\ref{eqn:otherTerms} and is a symmetry {\it of the likelihood at fixed $\mu$}.

In the more general case one can notice from Eqs.~\ref{eqn:comboProbInt1} and
\ref{eqn:comboProbInt2} that the probability densities integrated over $\mu$
are invariant to complex conjugation
of both ${\bf A} (t$) {\it and}  the overlap integral ${\cal I}_{mu}$ of Eq.~\ref{eqn:Imu}.  This
can be 
accomplished with a more complicated adjustment of $\delta_s$.  With a 
nonrelativistic Breit Wigner the required transformation is
\begin{displaymath}
  \delta_s \rightarrow 2\delta_{BW}-\delta_s
\end{displaymath}
where $\delta_{BW}\equiv \arg{\left(\log{\left((\mu_{hi}-\mu_{\phi}+i\Gamma_{\phi}/2)/(\mu_{lo}-\mu_{\phi}+i\Gamma_{\phi}/2)\right)}\right)}$.  The phase $\delta_{BW}$ reduces to $\delta_{BW}=0$ in the limit of an infinitesimally thin interval in $\mu$, and to $\delta_{BW}=-\pi/2$ in the limit of a finite symmetric interval. This demonstrates real differences in the two formulations,
and underscores the need for caution when applying the formulae of Ref.~\cite{ref:LHCb} to a finite interval in $\mu=m(K^+K^-)$.

\section {Conclusion}

In this paper we have presented a compact formalism to easily access physical observables in
the analysis of the decay $B^0_s\rightarrow J/\psi \phi$.  This formalism has practical applications,
since complex vectors and their vector-algebraic operations can be easily
implemented in high-level computer languages in order to model and generate such decays, but 
also because the symmetries of the formulae under operations such as rotation and complex conjugation are 
apparent and provide better physical insight into this complicated decay mode. The normalized probability densities  
can be used for the experimental extraction of physical parameters, in scenarios with no  $CP$ violation, with 
mixing-induced $CP$ violation, or even with direct $CP$ violation.  In case of mixing induced $CP$ violation, 
the effect of the $S$-wave contribution has also been included in the decay rate formulae.

\acknowledgments{We thank Barry Wicklund for extraordinarily valuable input, and
Yuehong Xie for helpful advice concerning the symmetries.  This work
was supported by the U.S.  Department of Energy and National Science
Foundation; the Bundesministerium f\"{u}r Bildung und Forschung, Germany;
the Science and Technology Facilities Council and the Royal Society,
UK; the Institut National de Physique Nucleaire et Physique des
Particules/ CNRS, France and the Comisi\'on Interministerial de Ciencia
y Tecnolog\'ia, Spain.}


\begin {thebibliography}{99}

\bibitem {ref:digheCKM} 
A. S. Dighe  et al. \epjc{6}{1999}{647} [\hepph{9804253}]

\bibitem {ref:LenzAndNierste} 
A. Lenz and U. Nierste, \jhep{0706}{2007}{72} [\hepph{0612167}]

\bibitem{ref:CDFUntagged}
T. Aaltonen et al. (CDF Collaboration), \prl{100}{2008}{121803} [\arXivid{0712.2348}]

\bibitem{ref:D0Untagged}
V. M. Abazov et al. (D0 Collaboration), \prl{98}{2007}{121801} [\hepex{0701012}]

\bibitem{ref:CDFTagged}
T. Aaltonen et al. (CDF Collaboration), \prl{100}{2008}{161802} [\arXivid{0712.2397}]

\bibitem{ref:D0Tagged}
V.M. Abazov et al. (D0 Collaboration), \prl{101}{2008}{241801} [\arXivid{0802.2255}]

%\bibitem{ref:CdfUnpublished} 
%The CDF Collaboration, /CDF/ANAL/BOTTOM/PUBLIC/10206 (unpublished)

%\bibitem{ref:D0Unpublished} 
%The D0 Collaboration, D0Note 6098-CONF (unpublished)

\bibitem {ref:InPursuit} 
I. Dunietz, R. Fleischer, and U. Nierste, \prd{63}{2001}{114015} [\hepph{0012219}]

\bibitem {ref:Sheldon} Sheldon Stone and Liming Zhang, \prd{79}{2009}{074024} [\arXivid{0812.2832}]

\bibitem {ref:dighe} 
A. S. Dighe et al. \plb{369} {1996} {144} [\hepph{9511363}]
 
\bibitem {ref:CK} 
M. Kobayashi and T. Maskawa, \ptp{49}{1973}{652}

\bibitem {ref:utfit} 
M. Bona et al. (UTfit Collaboration), \jhep {0610}{2006}{081} [\hepph{0606167}]

\bibitem {ref:CDFB0s}
T. Aaltonen et al. (CDF Collaboration), \prl{97}{2006}{242003} [\hepex{0609040}]

\bibitem {ref:wwerf}
W. Gautschi, \newjournal{SIAM J. Numer. Anal.}{SJNAA} {7}{1970}{187}

\bibitem {ref:GronauRosner} 
Michael Gronau and Jonathan Rosner, \plb{669}{2008}{321} [\arXivid{0808.3761}]

\bibitem {ref:LHCb} Y. Xie et al. \jhep{0909}{2009}{074} [\arXivid{0908.3627}]


\end {thebibliography}
\end{document}